\documentclass[preprint,12pt]{elsarticle}
\usepackage{amsmath}
\usepackage{amssymb}
\usepackage{multirow}
\usepackage{enumerate}
\usepackage{cases}
 \usepackage{amsthm}
 \usepackage{framed}

\usepackage{mathrsfs}

\makeatletter

\newcommand {\Rmnum} [1] {\expandafter \@slowromancap \romannumeral #1@}
\makeatother

\begin{document}

\begin{frontmatter}
\title{A voting scheme with post-quantum security based on physical laws}


\author{Hua Dong$^{1,2,3}$}
\author{Li Yang$^{1,2,3}$\corref{1}}
\cortext[1]{Corresponding author email: yangli@iie.ac.cn}
\address{1.State Key Laboratory of Information Security, Institute of Information Engineering, Chinese Academy of Sciences, Beijing 100093, China\\
2.Data Assurance and Communication Security Research Center,Chinese Academy of Sciences, Beijing {\rm 100093}, China\\
3.School of Cyber Security, University of Chinese Academy of Sciences, Beijing {\rm  100049}, China}

\begin{abstract}
Traditional cryptography is under huge threat along of the evolution of quantum information and computing. In this paper, we propose a new post-quantum voting scheme based on physical laws by using encrypted no-key protocol to transmit message in the channel, which ensures the post-quantum security. Unlike lattice-based and multivariate-based electronic voting schemes, whose security is based on the computational problems assumption that has not been solved by effective quantum algorithms until now, the security of the voting scheme based on the physical laws is depended on inherent limitations of quantum computers and not influenced by the evolution of new quantum algorithms. In detail, we also rigorously demonstrate that the scheme achieves the post-quantum security and all properties necessary for voting scheme such as the completeness, robustness, privacy, eligibility, unreusability, fairness, and verifiability.

\end{abstract}


\end{frontmatter}


\section{Introduction}

The voting scheme on the internet has been studied in recent decades. Since it is an attractive aspect of cryptography, there has been a lot of cryptographic voting schemes, which aim at achieving security and privacy simultaneously. Chaum proposed the first electronic voting scheme \cite{Chaum1981} in 1981. The scheme uses public key cryptography and pseudonyms rosters to conceal voters¡¯ identity, but does not ensure the privacy. Over the years, there have been many electronic voting schemes. Those schemes are divided into three categories: (1) The voting schemes based on homomorphic encryption   \cite{{Benaloh1987},{Cramer1996},{Benaloh1994},{Cramer1997}} (2) The voting schemes based on the Mix-net \cite{{ChaumThe1988},{Sako1995},{Park1994},{Michels1996}} (3) The voting schemes based on blind signature \cite{{Fujioka1993},{Ohkubo1999},{Chaum1977}}. However, so far, almost existing traditional voting schemes have been easily compromised by quantum algorithms\cite{Shor1997}, whose security assumption based on the classical assumption is integer factoring or discrete logarithm security assumption .

Therefore, proposing voting schemes based on the cryptographic algorithm that can resist quantum adversaries is an important issue. In order to achieve the goal, constructing a cryptographic algorithm that can resist quantum adversaries has become of general interest in recent years. The algorithms resisting quantum adversaries are divided into two categories as follows. One is based on quantum computing and quantum communication, belonging to quantum cryptography. There are some quantum voting schemes\cite{{Christandl2005},{Hillery2006},{Vaccaro2007},{Okamoto2008},{Bonanome2011},{Yang2012},{YangRui2013}} inspired by quantum cryptography. However, The overall system of quantum communication is not as complete as traditional communication. At present, it seems be more expensive and complicated than the traditional. The other is based on classical computing environment, which is as usual called post-quantum cryptography \cite{Bernstein2009}. There are some voting schemes \cite{{Sundar2014},{Chillotti2016},{Pino2017}} based on the hard problems based on lattice, multivariate linear equations and other computational problems. We have not found an effective quantum algorithm to solve the difficult problem of the above-mentioned post-quantum security cryptographic algorithm until now. We can call this ``passive defense" against quantum adversaries.
With the developing with quantum computing and algorithm, they may be compromised by new quantum algorithms proposed \cite{{Farhi2001},{Farhi2009},{Eldar2016}}. Therefore, facing the quantum adversaries, we have to consider defending them actively. Based on the above viewpoint, we can construct the voting schemes based on post-quantum cryptographic algorithms along another line. We can move the focus from passive defense to active defense, and study the characteristics of quantum computers and it's internal defects, which are based on the inherent physical laws. Since the quantum computer is a physical system, its gate operation rate is limited by some basic physical parameters. Thus, we can construct cryptographic algorithms to be post-quantum based on physical limitations. Some of these algorithms such as encrypted key exchange protocol (EKE) \cite{Bellovin1992} and encrypted no-key protocol(ENK) \cite{Yang2013} based on the above viewpoint were proposed , whose security is based on the physical laws which depends on inherent limitations of quantum computers. Due to the inherent physical laws, the security is not influenced by the evolution of new quantum algorithms.

On the above-mentioned viewpoint, we propose a voting scheme with post-quantum security based on physical laws, which is inspired by the ENK protocol with the post-quantum security \cite{Yang2013}. \cite{Yang2013} demonstrated the post-quantum security from respective of the physical laws, where the authors showed the lower limit of the time cost when the discrete algorithm of the ENK protocol is calculated for one cycle. Specifically, we use the ENK protocol to transmit message in the channel, which ensures the post-quantum security. And the message authentication code (MAC) \cite{Bellare1996} is introduced to prevent the messages from being tampered with by any party of our scheme and outsiders. With the help of administrator, voters can pass their ballots to counter anonymously in our scheme. Meanwhile, nobody can trace the ballots and match the voter's identity with the ballot. In addition, any party of our scheme can verify the validity of the ballot. These security properties are all based on the inherent physical laws of quantum computers, which are not relevant to the evolution of new quantum algorithm.

The rest of this paper is organized as follows. In the next section, we present the ENK protocol with post-quantum security and analyze its post-quantum security based on the physical laws. In Sect.3, we present our voting scheme in detail. Subsequently, we analyze the security of the scheme in Sect.4. Then we make a discussion about the aspects of the practical post-quantum security in Sect.5. Finally, we make a conclusion in Sect.6.
\section{Preliminaries}
In this section, we review the encrypted no-key protocol and its post-quantum security analysis based on physical laws, which will be used in the voting scheme.
\subsection{Encrypted No-key Protocol}

%
The encrypted no-key protocol \cite{Yang2013} will be used in our scheme to ensure the post-quantum security, which is developed from the Shamir no-key protocol\cite{Menezes1997}. In a no-key protocol, the sender and the receiver do not exchange any keys. However the protocol requires the sender and receiver to have two private keys for encrypting and decrypting messages. The following properties are required for the no-key protocol.

\begin{enumerate}
\item The algorithm in no-key protocol ia based on exponentiation modulo a large prime as both the encryption function $E(*)$ and decryption function $D(*)$. That is
\begin{align}\label{1}
                 E(e,m) & = m^e~mod~p, \\
               D(d,m) & = m^d~mod~p,
              \end{align}
 where $p$ is a large prime, $m$ is any message, $e$ is any encryption exponent and $d$ is the corresponding decryption exponent.
\item For any encryption exponent $e$ in the range $1..p-1$, there is
\begin{equation}\label{3}
 gcd(e,p-1)~=~1.
\end{equation}
\item  The corresponding decryption exponent $d$ is chosen such that
\begin{equation}\label{4}
de \equiv 1~(mod~p-1).
\end{equation}
It follows from Fermat's Little Theorem that
\begin{equation}\label{6}
D(d,E(e,m)) = m^{de} mod p = m.
\end{equation}
\item
The Shamir No-key protocol has the desired commutativity property since
\begin{equation}\label{5}
 E(a,E(b,m)) = m^{ab} mod p = m^{ba} mod p = E(b,E(a,m)).
\end{equation}
\end{enumerate}
\vspace{-0.5mm}
It is relatively easy to know that the Shamir no-key protocol does not ensure the post-quantum security and resist man-in-the-middle (MIM) attack. The \cite{Yang2013} proposed the ENK protocol, in which both parties pre-share a password $P$ before no-key communication, where $P$ is used for resisting the quantum adversaries and the MIM attack. The protocol is presented here.

\vspace{1.95mm}
\noindent\fbox{%
   \parbox{0.965\textwidth}{%
\vspace{1.95mm}
\textbf{~~Encrypted No-key Protocol}
\vspace{-0.75mm}
\begin{enumerate}
 \item Alice randomly chooses a message $M $ and a secret number $a$, then she
calculates $a^{-1}~(mod~q - 1)$ and sends ${E_{P}(M^{a}~mod~q)}$ to Bob;
     \item Bob randomly chooses a secret number $b$, decrypts with $P$ and sends ${E_{P}(M^{ab}~mod~q)}$ to Alice;
     \item Alice calculates $M^{b}~mod~q~=~((M^{a})^b)^{(a^{-1}~mod~q-1)}mod~q~$, and sends $E_{P}(M^{b}~mod~q)$ to Bob;
     \item Bob decrypts $E_{P}(M^{b}~mod~q)$ to recover M.
\end{enumerate}
}
}
\vspace{0.95mm}
\subsection{The Post-quantum Security of Encrypted No-key Protocol}

The post-quantum security of ENK protocol in ref.\cite{Yang2013} is specifically analyzed from the perspective of physical limitation. We know that the discrete logarithm (DL) is used in the ENK protocol. And if quantum adversaries use Shor algorithm to solve the DL problem, it requires a large number of controlled-NOT (CNOT) gate operations \cite{Michael2010}. The CNOT gate operations is limited by CNOT gate operation times and maximum number of operations for various candidate physical realizations of interacting systems of quantum bits. One is CNOT gate operation times and maximum number of operations. Since the qubits consisting of CNOT gates use phonons to interact with other collective excitation particles that are far apart from each other, the efficiency of quantum computer operations are limited by the movement speed of phonon or other medium. Therefore, there is a lower limit of the time cost when the discrete algorithm of the ENK protocol is calculated for one cycle, depending on the operating time of a single CNOT gate and the number of CNOT gates. The other is candidate physical realizations of quantum computers. If the quantum adversary Eve wants to know the messages which are transmitted by the ENK protocol, she must make a password-guessing attack. Whenever the adversary guesses the candidate password, she uses Shor algorithm to calculate the DL problem once. If the length of the password is $n$, she uses Shor algorithm to calculate the DL problem $2^n$ times. The total time required for the attack is so long for Eve that it is unrealistic.

Specifically, for the ENK protocol, the attacker based on some universal parameters of single-qubit quantum gate operations Eve wants to get the message of communication. She can do a password-guessing attack. For each guessing password $P'$, Eve should perform the discrete logarithm once. Let the lower bound of the time cost in the discrete logarithm calculation cycle be $\Delta T_{1}$, the time to perform a basic quantum logic operation be $\Delta t_{1}$ and the number of quantum gates serialized in the discrete logarithm algorithm be $N_{1}$.
  Then the time of a discrete logarithm computing cycle $\Delta T_{1}$ will be
  \begin{equation}\label{1}
    \Delta T_{1}=N_{1}\cdot \Delta t_{1}.
  \end{equation}
It is well known that $ \Delta t_ {1} $ has a lower bound:$\Delta t_{1}\geq 10^{-14}$. The value of $N_{1}$ determines the computational speed of the discrete logarithm algorithm in a quantum computer. From the ref.\cite{Yang2013}, the rough estimate of the lower bound is $10^{4}$, so we can get a discrete logarithm of the lower bound of time:
  \begin{equation}\label{1}
    \Delta T_{1}\geq 10^{4}\cdot \Delta t_{1}\geq 10^{4}\cdot10^{-14}=10^{-10}.
  \end{equation}
In the real physical world, we consider that the continuous attack duration is $2^{32}$(100 years). Let the number of times that the quantum adversary needs to crack the password within the effective attack duration be $N$, we have
  \begin{equation}\label{1}
    N<\frac{2^{32}}{10^{-10}}<2^{66}.
  \end{equation}
  For each candidate $P'$, the length of the password P should satisfy $|P|\geq 66$. That is, for resisting attack with several quantum computers, a 68-bit password is enough to ensure the security of an ENK within the effective attack duration in the real physical world.

The above analysis is based on some common parameters of the single qubit gate operation. Compared with the ion-trap quantum computer\cite{Cirac1995}, it is easy to know that the computational power of the adversary using an ion trap computer is stronger than that of a single qubit gate operation based on common parameters. The ion-trap is one of the earliest implementations of quantum computer and has a series of advancements in implementing the Shor algorithm. In recent years, it is considered to be one of the most promising physical implementations of quantum computer. For adversaries with an ion-trap computer, she also do a password-guessing attack. For each guessing password $P'$, Eve should perform the discrete logarithm once. Let the the lower bound of time finishing a discrete logarithm computation be $\Delta T_{2}$, the time that a CNOT operation performs be $\Delta t_{2}$ and the number of CNOT operations performing serially bek $N_{2}$. By analyzing the relationship between frequency, the wavelength of the acoustic wave and the mass of every ion, we can have
  $\Delta t_{2}~\approx~2.85\times10^{-4}$. Due to ref.\cite{Cleve2000}, we can get $N_{2}~\sim~(\log n)$. In view of the fault-tolerant structure of ion trap quantum computers, especially the error correction coding related to the threshold theorem of concatenated quantum, we know that
\begin{equation}\label{1}
  N_{2}~>~10^{2},
\end{equation}
then we have
\begin{equation}
\begin{split}
    \Delta T_{2} = N_{2}\cdot \Delta t_{2} \geq 2.85 \times 10^{-2},
\end{split}
\end{equation}
Although the physics parameters other types of quantum computer are different, the conclusion is similar. In the real physical world, we consider that the continuous attack duration is $2^{32}$(100 years). Within one second, the ion trap quantum computer can not do the discrete logarithm calculation $2^{6}$, so the attacker can not perform the discrete logarithm calculation $2^{38}$ times within the effective attack duration. Assuming that the size of a quantum computer is about one square meter, the upper limit of the number of quantum computers used by any adversary is
\begin{equation}\label{1}
  4\pi\times(6370\times 10^{3})^{2}~=~5.1\times 10^{14}~<~2^{49}.
\end{equation}
Let the number of times that the quantum adversary needs to crack the password within the effective attack duration be $N$, we have
 \begin{equation}\label{1}
   N<2^{87}.
  \end{equation}
When the length of password $P$ is 88, the quantum adversary must solve the algorithm $2^{87}$ times on average, which is beyond the maximum computational power of the attacker within the effective attack duration for these quantum computers. For more detailed argument of the security, we refer the readers to \cite{Yang2013}.

\section{The Voting Scheme}
\subsection{Notations of the Voting Scheme}
The roles in the scheme are voters $V_{i}~(1\leq i\leq n)$, administrator $A$ and counter $C$.  With the help of administrator, voters can pass their ballots to counter anonymously. The notations involved in the scheme are described in table.1.
\begin{table}[!htbp]
\linespread{1.15}\selectfont
\scriptsize
\begin{tabular}{@{}p{.1\textwidth}p{.85\textwidth}}
\hline
\small{Notation}                         &\small{Description}
\vspace{0.5mm}                                      \\\hline
$\parallel$            &       \small{concatenation of two bit strings }                   \\
\small{$\mathcal{B}_{i}$}                         &\small{The ballot of $V_{i}$ }                 \\
\small{$\ell$}                       &\small{Candidate set             }                 \\
\small{$V_{i}$}                           & \small{Voter$_{i}$ who has legal voting right }         \\
\small{$ID_{i}$}                  & \small{Identification number string of $V_{i}$    }                      \\
\small{$ID_{j}$}                  & \small{Replaced identification number string of $V_{i}$   }                 \\
\small{$S_{i}$} & \small{a unique verification string corresponding to $ID_{i}$} \\
\small{${P_{av_{i}}}$}                            &\small{ The password of $V_{i}$ and $A$'s ENK protocol  }   \\
\small{$K_{va}$}      & \small{The key shared by all voters and $A$ }\\
\small{$K_{a}$ }        & \small{an authentication key of $A$ and $C$}\\
\small{$P_{ac}$ }                      & \small{The password of $A$ and $C$'s ENK protocol}\\
\small{$K_{vc}$}                            & \small{The key shared by all voters and $C$  } \\
\small{${P_{vc}}$}                        & \small{The password of all voters and $C$'s ENK protocol  }        \\
\small{$a_{{cv}_{i}}$}      & \small{The random key generated by $V_{i}$ performing ENK protoocl with $C$}\\
\small{$b_{{cv}_{i}}$}      & \small{The random key generated by $C$ performing ENK protoocl with $V_{i}$}\\
\small{$h_{K_{\centerdot}}(\centerdot)$}       &  \small{The MAC of ($\centerdot$) encrypted with $K_{\centerdot}$} \\
\small{$X_{i}$}                        & \small{      $\mathcal{B}_{i}\|S_{i}\|h_{K_{va}}(\mathcal{B}_{i}\|S_{i})$ }\\
\small{$Y_{i}$}                        &  \small{   $X_{i}\|h_{K_{vc}}(X_{i})$              }\\
\small{$E^{*}_{K_{\centerdot}}[\centerdot]$}&\small{The Symmetric encryption algorithm with $K_{(\centerdot)}$  } \\
\small{$E_{P_{\centerdot}}[\centerdot]$}           &\small{ The ENK protocol with $P_{(\centerdot)}$}\\
\hline
\end{tabular}
\caption{Notations of the voting scheme}
\end{table}
\subsection{Construction of the Voting Scheme}
We present the voting scheme in detail in this section. First, administrator $A$ publishes a candidate set, distributes an $ID$ for the voter to show the legal identity and pre-processes the keys required for the voting. Then voter sends the voting request to $A$, which contains voter's $ID$ and encrypted ballot. After authentication, $A$ helps the authenticated voter to pass the ballot to counter $C$ using the ENK protocol. Among them, the ballot information is encrypted by the double-encryption and voter'$ID$ is replaced by another $ID$, so as to ensure the security of the election. Finally, the ballots are counted and the results are published by $C$. The model of of the voting scheme is described in the following Figure.1.

\begin{figure}[htbp]
  \centering
  \centering\includegraphics[width=5in]{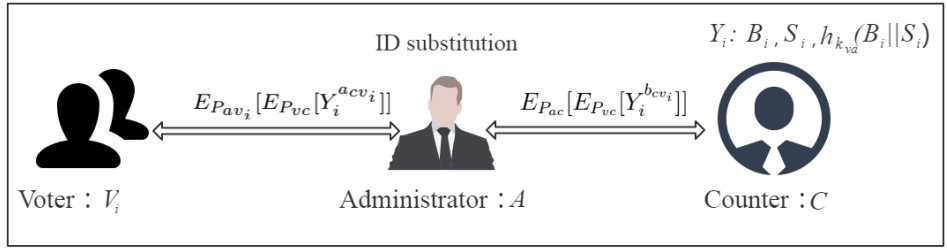}
  \caption{The Flow diagram of the voting scheme. We can take voter $V_{i}~(1\leq i\leq n)$ as an example. With the help of administrator, voter can pass his encrypted ballot information, $Y_{i}$, to counter anonymously, where $Y_{i}$ consist of $\mathcal{B}_{i},S_{i}$ and $h_{K_{va}}(\mathcal{B}_{i}\|S_{i})$.  Then $C$ records it and announces the voting result in final. The notations' description of the model is shown in detail in Table.1 on the above. }\label{fig:digit}
\end{figure}

 The voting scheme consists of initial phase, authentication and voting phase, delivering ballot phase and publishing ballot phase. The structure of of the voting scheme is following.

\begin{table}[h]
\vspace{-3mm}
\begin{tabular}{ll}	
			Initial phase: &Candidates are opened; Keys are preprocessed.\\
			Authentication and voting:&Voters pass authentication and start voting.\\
		    Delivering ballot phase :&$A$ helps voters deliver ballots.\\
			Publishing ballot phase:&$C$ publishes the voting result.\\
\end{tabular}
\vspace{-3mm}
\end{table}

We use the ENK protocol to pass ballot information to ensure the post-quantum security. To prevent ballot information from being tampered, we use the unconditionally secure MAC. There are a lot of unconditionally secure MACs; for instance, \cite{Krawczyk1994}. We do not specify which MAC is exact to be used for the scheme. We use ID substitution operation to ensure the privacy of voters. Because of this operation, even if $C$ gets ID, he does not know the exact corresponding voting identity. The specific scheme is as follows:

\noindent\textbf{Phase \uppercase\expandafter{\romannumeral1}}:\textbf{ Initial}

In the initial phase, firstly voting candidate set is announced. Then the communication keys required for the voting scheme are pre-distributed and the communication identity strings about voters and $A$ are pre-shared. The steps in initial phase are as follows:
\begin{enumerate}
  \item The voting candidate set is announced.

  Supposing there are $m$ candidates, there is a candidate set.
  \begin{equation*}
  \ell=\{\mathcal{B}^{1},\mathcal{B}^{2},\mathcal{B}^{3},...,\mathcal{B}^{m}\}.
  \end{equation*}
  Each element of the set is an s-bit string that represents an eligible candidate, i.e. $\mathcal{B}^{j}\subseteq \{0,1\}^{s}(j\in[1,m])$. The administrator announces the set $\ell$ and each candidate corresponding to the $s$-bit string. We assume s is large enough to ensure that the probability is negligible where a random $s$-bit string is an element of set $\ell$. For an eligible voter $V_{i}$, he chooses one candidate as his ballot $\mathcal{B}_{i}$.

  \item The necessary preparations for voting are completed.

 (1). Preparations between voters and administrator

    Each eligible voter $V_{i} (1\leq i\leq n)$ has a bunch of numbers $ID_{i}$ that represents voter's identity, which is distributed by $A$. Each voter $V_{i}$ and $A$ pre-distribute a password $P_{{av}_{i}}$ , which is used for the ENK protocol. In addition, all voters and $A$ pre-distribute a communication key $K_{va}$, which is used to encrypt voter's $ID$. These are shown in Table.2.

 (2). Preparations between administrator and counter

  $A$ and $C$ pre-distribute a password $P_{ac}$ , which is used for the ENK protocol. In addition, they also pre-distribute a communication key $K_{ac}$, which is used to encrypt voter's $ID$. These are also shown in Table.2.

  (3).Preparations between voters and counter

  All voters and $C$ pre-distribute a password $P_{vc}$ , which is used for the ENK protocol. In addition, they also pre-distribute a communication key $K_{vc}$, which is used to encrypt voter's $ID$. These are also shown in Table.2.

\vspace{-2mm}
\begin{table}[htbp]
  \centering
  \begin{tabular}{|c|c|}\hline
			Both sides of communication&\multicolumn{1}{c|}{$~~$Preprocessed key information}\\\hline
			
			$V_{i}~~ \&~~ A$&$K_{va},~P_{{av}_{i}}$\\
			
		    $A ~~\& ~~C$&$K_{ac},~P_{ac}$\\
			
			$V_{i} ~~\&~~ C$&$K_{vc},~P_{vc}$\\\hline
		\end{tabular}
		\caption{The preprocessed keys }
		\label{tab:Margin_settings}
\end{table}
\vspace{-6mm}
\end{enumerate}

\noindent\textbf{Phase \uppercase\expandafter{\romannumeral2}}:\textbf{ Authentication and Voting }

In this phase, voters who want to vote should be authenticated. If $V_{i}$ is a legal voter, he can pass the authentication with a valid $ID$ and transmit the encrypted ballot information to $A$, where the encrypted ballot information can prevent from being tampered with; $A$ stores $V_{i}$' information into the database and gets a voters' information table passing authentication. We take the specific voter $V_{i}$ as an example, then the steps in authentication and voting phase are as follows:

\begin{enumerate}
  \item  $V_{i}$ chooses his ballot and generates the unique verification string corresponding to $ID_{i}$.

  $V_{i}$ chooses $\mathcal{B}_{i}$ as his own ballot and generates a unique verification string $S_{i}$.
  where $S_{i}$ is the unique verification string corresponding to $ID_{i}$. In addition, $|S_{i}|$ is large enough to ensure that the probability of generating a same string is negligible.
  \item  $V_{i}$ generates encrypted ballot information.

    To prevent $\mathcal{B}_{i}\|S_{i}$ from being tampered, $V_{i}$ generates a message authentication code for $\mathcal{B}_{i}\|S_{i}$ using $K_{va}$, where $K_{va}$ is the key shared by all voters and $A$. To simplify the following, we do the definition.
                  \begin{equation}\label{1}
                   X_{i} \triangleq \mathcal{B}_{i}\|S_{i}\|h_{K_{va}}(\mathcal{B}_{i}\|S_{i}),
                  \end{equation}
                  \begin{equation}\label{1}
                  Y_{i} \triangleq X_{i}\|h_{K_{vc}}(X_{i}) ,
                  \end{equation}
  where $h_{K_{va}}(\mathcal{B}_{i}\|S_{i})$ is the MAC of $\mathcal{B}_{i}\|S_{i}$ with the key $K_{va}$ to prevent from being tampered with, $K_{vc}$ is the key shared by all voters and $C$ and $h_{K_{vc}}(X_{i})$ is the MAC of $X_{i}$.

 Each voter $V_{i}$' encrypted ballot information is double-encrypted as follows:
\begin{center}
$ E_{P_{{av}_{i}}}[E_{P_{vc}}[Y_{i}^{a_{{cv}_{i}}}], $
\end{center}

    where, the inner layer is encrypted by $P_{vc}$, which is used for the ENK communication between each voter and $C$ and the outer layer is encrypted by $P_{av_{i}}$, which is used for the ENK communication between each voter $V_{i}$ and $C$. Due to the inner layer encrypted by $P_{vc}$ , it ensures that $A$ is invisible for the encrypted ballot information.
  \item $V_{i}$ transmits the encrypted ballot information to $A$ and finishes the authentication.

 $V_{i}$ transmits the encrypted ballot information to $A$.

  \begin{center}
  $V_{i}\xrightarrow{(E^{*}_{K_{va}}[ID_{i}],~ E_{P_{{av}_{i}}}[E_{P_{vc}}[Y_{i}^{a_{{cv}_{i}}}]])} A$,
  \end{center}
   where $ID_{i}$ is the legal voter¡¯s identification number string, $E^*$ is the Symmetric encryption algorithm with $K(\bullet)$, $P_{vc}$ is the password of all voters and $C$¡¯s ENK protocol, $P_{{av}_{i}}$ is the password of $V_{i}$ and $A$¡¯s ENK protocol, $E_{P_{vc}}[\ast]$ is executing the ENK protocol with $P_{vc}$ and $E_{P_{{av}_{i}}}$ is executing the ENK protocol with $P_{{av}_{i}}$.

   Then $A$ decrypts $E^{*}_{K_{va}}[ID_{i}]$ with $K_{va}$ to get $ID_{i}$. If $ID_{i}$ exits in the database, $A$ refuses to deliver the ballot to prevent $V_{i}$ from repeating voting. Otherwise, $V_{i}$ chooses the corresponding $P_{{av}_{i}}$ according to $ID_{i}$ to execute the ENK protocol with $A$. If $ID_{i}$ is not eligible, $V_{i}$ does not pass authentication.
   \item $A$ stores $V_{i}$' information into the database.

   $A$ puts the legal voters' information ($ID_{i},l,V_{i}$) into the database, where $l$ is the entry of voter authentication. After authenticating all legal voters, $A$ gets a voters' information table passing authentication as shown in Table.3 and announces the $ID$ of the authenticated voters. So that voters know that they have passed authentication.
\end{enumerate}

\begin{table}[htbp]
  \centering
  \begin{tabular}{|c|c|c|c|}\hline
			Entry&\multicolumn{2}{c|}{Information of voters}\\\hline
            &identity sequence&Voter's identity\\
			1&$ID_{i}$&$V_{i}$\\
			$\vdots$&$\vdots$&$\vdots$\\
		  $l$&$ID_{k}$&$V_{k}$\\
			$\vdots$&$\vdots$&$\vdots$\\
			n&$ID_{t}$&$V_{t}$\\\hline
		\end{tabular}
		\caption{voters' information table passing authentication}
		\label{tab:Margin_settings}
\end{table}

\noindent\textbf{Phase \uppercase\expandafter{\romannumeral3}}:\textbf{ Delivering ballot}

In this phase, $A$ helps voters to deliver their ballots anonymously. At first, $A$ performs an ID substitution operation one by one, which is used to hide the identity of the voters. Next, $A$ sends the ballot information with replaced $ID$ to $C$. Finally the ballot is delivered to $C$ through the middleman $A$. We also take $V_{i}$ as an example, then the steps in delivering ballot phase are as follows:
%
\begin{enumerate}
  \item  $A$ performs an ID substitution operation.

  When $A$ helps $V_{i}$ to deliver ballot, the $ID_{i}$ of $V_{i}$ is replaced by $ID_{j}$ and the ballot is delivered in the $j$-th order. At this point, only $A$ knows ID substitution table, even if $C$ sees $ID$ , $C$ does not know the specific corresponding voter identity. The ID substitution table as shown in Table 4.
  \begin{table}[htbp]
  \centering
  \begin{tabular}{|c|c|c|}\hline
			Identity sequence&\multicolumn{2}{c|}{Relevant information after replacement}\\\hline
            &Entry&Replaced identity sequence\\
			$ID_{k}$&1&$ID_{1}$\\
			$\vdots$&$\vdots$&$\vdots$\\
		  $ID_{i}$&$j$&$ID_{j}$\\
			$\vdots$&$\vdots$&$\vdots$\\
			$ID_{t}$&$n$&$ID_{n}$\\\hline
		\end{tabular}
		\caption{ID substitution table}
		\label{tab:Margin_settings}
\end{table}
  \item The ballot is delivered to $C$ through the middleman $A$.

  In this step, $A$ delivers the encrypted ballot information to $C$. After substitution, $V_{i}$' encrypted ballot matches with $ID_{j}$. Due to double encryption, $A$ cannot transmit the encrypted to C through a round of ENK communication. $A$ acts as an intermediary to create anonymous interactive communications between voters and $C$.

  In these anonymous interactive communications, double encryption of the $V_{i}$' encrypted ballot information changes. The outer layer is changing with the both sides of communication. When the both sides of communication are $A$ and $C$, the outer layer is constant that is encrypted by $P_{vc}$ using the ENK communication between $A$ and $C$. When the both sides of communication are $V_{i}$ and $C$, the outer layer is replaced by $P_{av_{i}}$ using the ENK communication between $V_{i}$ and $C$. In addition, the inner layer encrypted by $P_{vc}$ is fixed. These anonymous interactive communications is as follows:
\begin{center}
                  $A\rightarrow C:$ $(E^{*}_{K_{ac}}[ID_{j}],E_{P_{ac}}[E_{P_{vc}}[Y_{i}^{a_{{cv}_{i}}}]])$ ;

                $C\rightarrow A:$ $(E^{*}_{K_{ac}}[ID_{j}],E_{P_{ac}}[E_{P_{vc}}[Y_{i}^{a_{{cv}_{i}}b_{{cv}_{i}}}]])$ ;

               $A\rightarrow V_{i}:$ $((E^{*}_{K_{va}}[ID_{i}],E_{P_{av_{i}}}[E_{P_{vc}}[Y_{i}^{a_{{cv}_{i}}b_{{cv}_{i}}}]])$ ;

               $V_{i}\rightarrow A:$ $((E^{*}_{K_{va}}[ID_{i}],E_{P_{av_{i}}}[E_{P_{vc}}[Y_{i}^{b_{{cv}_{i}}}]])$ ;

               $A \rightarrow C:$ $((E^{*}_{K_{ac}}[ID_{j}],E_{P_{ac}}[E_{P_{vc}}[Y_{i}^{b_{{cv}_{i}}}]])$ ;
\end{center}
 where, these anonymous interactive Cs' principles are similar. We take ($E^{*}_{K_{ac}}[ID_{j}],E_{P_{ac}}[E_{P_{vc}}[Y_{i}^{a_{{cv}_{i}}}]])$ as an example between communication $A$ and $C$. Where $E_{K_{ac}}(*)$ is a symmetric encryption algorithm with encryption and decryption key $K_{ac}$. Then $C$ decrypts with $K_{ac}$ and gets $V_{i}$'s replaced identity string $ID_{j}$. Thus, $C$ checks whether voters repeat voting. After that $A$ and $C$ use the password $P_{ac}$ and execute the ENK protocol to transfer the $V_{i}$' encrypted ballot information. When $C$ gets $Y_{i}^{a_{{cv}_{i}}}$, $C$ adds $b_{{cv}_{i}}$ to the $V_{i}$' encrypted ballot information to encrypt the inner layer as the ENK protocol in Sect.2. Then the information that has been attached to $b_{{cv}_{i}}$ is sent to $A$ by $C$. The rest communications are similar to it.
\end{enumerate}

\noindent\textbf{Phase \uppercase\expandafter{\romannumeral4}}:\textbf{ Publishing ballot}

 In this phase, $C$ announces the valid ballot received and the voting result. First, $C$ checks whether the ballot is legal and the ballot has been tampered with. Every time a valid ballot is received by $C$, it is recorded on the bulletin board. After $C$ receives all ballots, $C$ will announce the bulletin board and the voting result. The publishing ballot phase is implementing as follows:

 \begin{enumerate}
   \item $C$ checks the validity of the ballot.

   We can take $V_{i}$ as an example. $C$ receives $Y_{i}$ from $V_{i}$, which consisted of $X_{i}$ and $h_{k_{vc}}(X_{i})$. Meanwhile, $C$ could receive replaced ID sequence $ID_{j}$. Firstly, $C$ uses $K_{vc}$ to reconstruct $h^{'}_{k_{vc}}(X_{i})$, which is the reconstructed MAC. If $h^{'}_{k_{vc}}(X_{i})$ isn't equal to $h_{k_{vc}}(X_{i})$, it proves that the message has been tampered. Otherwise, $C$ extracts $h_{K_{va}}(\mathcal{B}_{i}\|S_{i})$. Then $C$ extracts $\mathcal{B}_{i}$ and $S_{i}$. The next is that $C$ verifies if $S_{i}$ has been received. After the verification, if $\mathcal{B}_{i}\in \mathcal{L}$, $C$ considers the ballot valid and counts it. Finally, $C$ records $\mathcal{B}_{i},S_{i},h_{K_{va}}(\mathcal{B}_{i}\|S_{i})$ into Publishing ballot information table.

   \item $C$ will announce the bulletin board and the voting result.

 Every time a valid ballot is received by $C$, it is recorded on the bulletin board shown as Table 5.
  After all the votes of $n$ voters have been dealt with, $C$ counts and publishes the voter's verification string and the success of the ballot result. In addition, $C$ also publishes the replaced ID number which is replaced by $A$ for voters who did not vote successfully,. Thus, both $A$ and $V_{i}$ can know whether $V_{i}$ voted successfully. For each voter who did not vote successfully, $A$ creates a new series of keys £¨$ID^{'}_{j},P^{'}_{av_{i}}$£© and helps them to make a new round of voting.
When all the ballots are counted, announces the voting results and $\mathcal{B}_{i},S_{i},h_{K_{va}}(\mathcal{B}_{i}\|S_{i})$. Through the MAC of ballot and string $h_{K_{va}}(\mathcal{B}_{i}\|S_{i})$, voters can check if their votes are counted correctly and administrator can supervise $C$ to prevent ballots from being tampered with.
 \end{enumerate}

\begin{table}[htbp]
  \centering
  \begin{tabular}{|c|c|c|c|}\hline
		Entry&Ballot &Verification string&MAC of ballot and string\\\hline
		 1 & $\mathcal{B}_{k}$&$S_{k}$&$h_{K_{va}}(\mathcal{B}_{k}\|S_{k})$\\
	 $2$& $\mathcal{B}_{i}$&$S_{i}$&$h_{K_{va}}(\mathcal{B}_{i}\|S_{i})$\\
		$\vdots$&	$\vdots$&$\vdots$&$\vdots$\\
		 $n$& $\mathcal{B}_{t}$&$S_{t}$&$h_{K_{va}}(\mathcal{B}_{t}\|S_{t})$\\\hline
		\end{tabular}
		\caption{Publishing ballot information}
		\label{tab:Margin_settings}
\end{table}
\section{Security analysis}
In this section, we are to discuss the security properties of the voting scheme. The voting scheme based on the physical laws reaches the post-quantum security and voting scheme criteria. The unreusability can be ensured, because each party of the voting scheme has his own recorded database to prevent replay attacks in each phase. Meanwhile, the eligibility can be ensured, because one party needs to confirm the other legal identity before communication. In addition, our scheme has universal verification, because all ballots and voting result can be verified by three parties of the scheme. The above properties are easier to prove. Meanwhile, there is no mention of security property, such as privacy and robustness. In our scheme they are promised by the security of the ENK protocol based on physical laws, so we also focus on analyzing them.  In addition, we also demonstrate that the scheme achieves the others properties necessary for voting scheme in detail.

\noindent\textbf{Completeness.}
The completeness means that all the valid ballots must be counted correctly when all parties of the scheme are honest. The completeness is obviously satisfied if the voters, the administrator $A$ and counter $C$ execute the scheme honestly.


\noindent\textbf{Robustness.}
The robustness means that the dishonest parties of the scheme or outsiders cannot disrupt the voting scheme. The abnormal behaviors will be found, including communication terminated between any two parties and invalid messages delivered. We demonstrate the robustness when one party of the scheme or outsider wants to disrupt the scheme.

 When a voter $V_{e}$ is dishonest, there will be the following two cases: refusing to communicating with $A$ or $C$ and sending an invalid ballot. In the first case, $A$ and $C$ do not think it is normal that the numbers of voters voting in the scheme is less than the total number of voters. For example, during the authentication and voting phase, the voter $V_ {e}$ refuses to communicate with $A$ after obtaining the password $ {P_ {av}}_{i}$. $A_{w}$ could examine the communication numbers that obtains the password before the next phase. Thus, $A_{w}$ will find out the abnormal behavior of the voters $ V_ {i} $. In the second case, $V_{e}$ sends an invalid ballot $\mathcal{B}_{e}(\mathcal{B}_{e}\not\subseteq\mathcal{B})$ to $C$. Since $C$ verifies the legality of the ballot information $\mathcal{B}_{e}$ before recording the ballot. If the ballot information is valid ($ \mathcal{B}_{e} \in \ell $), $C$ considers that the ballot is valid and counts the ballot $ \mathcal{B}_{e} $. If not, she can refuse to receive the ballot from the dishonest voter and $V_{e}$' invalid ballot is not recorded in the bulletin board.

When a administrator $A_{e}$ is dishonest, there will be the following two cases: refusing to communicate with $C$ or sending changed encrypted ballot information. In the first case, similarly, $C$ can find the abnormal behavior if she does not receive the messages from $A_{e}$. In the second case, $A_{e}$ sends changed encrypted ballot information to $C$, not original message. Because $A_{e}$ does not know the key $K_{vc}$, which is shared by all voters and $C$. She do not create a valid message due to the unforgeability of MAC. After the decryption, $C$ gets a random string. The probability of randomly generating a valid string is negligible because the length of $S_{i}$ is so big. Therefore, $A_{e}$'s abnormal behavior is found out.

Then, when a counter $C_{e}$ is dishonest, She makes trouble on the bulletin board which consists of $\mathcal{B}_{i},S_{i},h_{K_{va}}(\mathcal{B}_{i}\|S_{i})(1 \leq i \leq n)$ . Due to $h_{K_{va}}(\mathcal{B}_{i})$, administrator $A$ can supervise $C$ to prevent ballots from being tampered with.

Finally, the outsiders who want to disrupt the scheme terminate the communication
between any two parties of our scheme or tamper with the encrypted ballot information. It is similar to the above mentioned analysis. So, the outsiders' abnormal behavior also is found out.

\noindent\textbf{Privacy.}
The privacy means that the content of the ballot is invisible to others except for the voter and $C$. In other words, the content and the voter's identity cannot be matched. In our voting scheme, we ensure the privacy based on the physical laws, which can resist the quantum adversaries.

In this scheme it can be assumed that administrator $A$ and counter $C$ are independent parties, i.e., they will not collaborate on tracking ballots.

First, we briefly discuss that the participants of our scheme are dishonest as the following two cases. One case is that privacy still exists when the administrator $A$ is dishonest. It is clear that $A$  knows the identity of voters but can not see the ballot due to the double encryption. The other case is that that counter $C_{e}$ is dishonest. $C_{e}$ can only know $\mathcal{B}_{i}\|h_{K_{va}}(\mathcal{B}_{i})$, but does not know the sender's identity. She can not match $\mathcal{B}_{i}$ with $ V_ {i} $'s identity.

Next, we demonstrate that when there exits an outsider Eve, the security of privacy is post-quantum secure. Any attacker wants to track ballots, which breaks the privacy. However, the privacy of our scheme based on the physical laws is equal to the security of the ENK protocol, since any information about voter identity and ballot is transmitted through the ENK protocol in the channel. Because the ENK protocol has post-quantum security, the privacy of our scheme is post-quantum. Specifically, If an attacker wants to track the ballot and break the privacy of our scheme, she must decrypt the communication in the channel encrypted with $ P_ {av_{i}} $, $ P_ {ca}$, and $ P_ {cv} $. Because each voter $V_i$¡¯ encrypted ballot information is double-encrypted, we can take the outer layer of double encryption in communication between $A$ and $C$ as an example. If Eve wants to get the messages of the communication, $E_{P_{ac}}[E_{P_{vc}}[Y_{i}^{a_{{cv}_{i}}}]]$ , she can conduct key-guessing attack ($a_{ac}$ and $b_{ac}$ are random numbers generated by the communication parties in the ENK protocol in Sect.2):

%
%
\begin{enumerate}
\item Eve chooses a candidate password and decrypts the messages.

Eve randomly generates a candidate password $P_{ac}^{'}$£¬then uses it to decrypt the messages of the channel and obtains
\begin{center}
 $((E_{P_{vc}}[Y_{i}^{a_{{cv}_{i}}}])^{a_{ac}})^{'}$, $((E_{P_{vc}}[Y_{i}^{a_{{cv}_{i}}}])^{b_{ac}})^{'}$ and $((E_{P_{vc}}[Y_{i}^{a_{{cv}_{i}}}])^{a_{ac}b_{ac}})^{'}$;
\end{center}
\item Eve extracts the random numbers generated by the communication parties.

Then he uses $((E_{P_{vc}}[Y_{i}^{a_{{cv}_{i}}}])^{b_{ac}})^{'}$,  $((E_{P_{vc}}[Y_{i}^{a_{{cv}_{i}}}])^{a_{ac}b_{ac}})^{'}$ to extract $a_{ac}^{'}$£¬
and $((E_{P_{vc}}[Y_{i}^{a_{{cv}_{i}}}])^{b_{ac}})^{'}$, $((E_{P_{vc}}[Y_{i}^{a_{{cv}_{i}}}])^{a_{ac}b_{ac}})^{'}$ to extract $b_{ac}^{'}$;
\item Eve gets the final messages after decryption.

Finally, he calculates $((((E_{P_{vc}}[Y_{i}^{a_{{cv}_{i}}}])^{'})^{a_{ac}^{'}})^{-1}$ and $((((E_{P_{vc}}[Y_{i}^{a_{{cv}_{i}}}])^{b_{ac}})^{'})^{b_{ac}^{'}})^{-1}$
   \end{enumerate}

He verifies whether the candidate password $P_{ci}^{'}$ is correct by checking whether $((((E_{P_{vc}}[Y_{i}^{a_{{cv}_{i}}}])^{a_{ac}})^{'})^{a_{ac}^{'}})^{-1}$ is equal to $((((E_{P_{vc}}[Y_{i}^{a_{{cv}_{i}}}])^{b_{ac}})^{'})^{b_{ac}^{'}})^{-1}$. It is obvious that the computational complexity of this attack depends on the length of password $P_{ac}$ . For each candidate $P_{ac}^{'}$, Eve has to solve the discrete logarithm problem. As the ENK protocol with post-quantum security in Sect.2, the quantum adversary must solve the algorithm $2^{87}$ times on average when the length of password P is 88, which is beyond the maximum computational power of the attacker within the effective attack duration.

\noindent\textbf{Eligibility.}
Eligibility means that only eligible voters are allowed to vote.
As stated in the authentication and voting phase, the identity of the voter $ V_ {i}(1\leq i\leq n)$ who applied for the voting are verified by administrator. Only eligible voters can be authenticated to access the delivering ballot phase. If an ineligible voter $V_{e}$ wants to vote successfully, she impersonates an eligible voter $V_{i}$ in the authentication and voting phase. However $V_{e}$ doesn't get the valid $ID_{i}$ and the communication key by eavesdropping the channel due to the ENK protocol with post-quantum security.

\noindent\textbf{Unreusability.} Unreusability means that each eligible voter cannot vote successfully twice.
In our scheme, each party of the voting scheme has his own recorded database to prevent replay attacks in each phase. Specifically, in the authentication and voting phase, $A$ can verify whether voters apply repeatedly based on the voters¡¯ information table passing authentication, namely Table.3. In the publishing ballot phase, $C$ verifies whether they have received repeated votes from the same voter by $S_{i}$ of publishing ballot information table, namely Table.5. As a result, it is impossible for each eligible voter who holds the voting right to repeat voting.

\noindent\textbf{Fairness.} Fairness means that nothing can affect the voting, especially that the counting of ballots does not affect the voting.
That means that the ballot information will not be leaked until the result is published. In the scheme, the authentication and voting phase is done after the authentication phase, and $C$ will not disclose the intermediate result of the voting scheme to others before the whole scheme is completed. Therefore the previous voters will not affect the subsequent voters. The scheme is fair for all voters.

\noindent\textbf{Verifiability.} Verifiability means that an eligible voter can verify that his ballot has been correctly counted or not. In addition, our scheme ensures universal verifiability, which shows all ballots and voting result can be verified by three parties of the scheme. In publishing ballot phase, $C$ publishes the publishing ballot information table which consists of $\mathcal{B}_{i},S_{i}$ and $h_{K_{va}}(\mathcal{B}_{i}\|S_{i})~(1\leq i \leq n)$. $V_{i}$ can locate his own verification string $S_{i}$ in Table.5 and verify the corresponding ballot information is correct or not. All voters and $A$ who own $K_{va}$ could check the validity of the ballot by $H_{K_{va}}(\mathcal{B}_{i}\|S_{i})$. $C$ can check ballots' validity by verifying whether $\mathcal{B}_{i}\in \mathcal{B}$. Therefore all participates in the scheme can check whether the voting result is normally counted.

\section{Discussion}
The post-quantum security of the proposed scheme that can resist quantum computers is based on the physical laws. In the voting scheme, we use the ENK protocol to transmit messages in the channel. From the \cite{Yang2013}, the ENK protocol with post-quantum security is to resist the quantum computer driven by typical coherent fields based on the ion-trap. From the \cite{Cirac1995}, it mainly analyzes the permitted logic depth of quantum computer driven by coherent fields. From the \cite{Yang2013}, the authors specifically analyze whether the quantum algorithm is able to perform reliably if the logical depth of the quantum algorithm is within the permitted logic depth. In our voting scheme, we use the ENK protocol with post-quantum security based on the physical laws, which cannot be attacked in the ion-trap quantum computing environment. In future, we can move on to extend the scope of our research to explore the security of voting scheme in quantum computing environments such as cavity quantum electrodynamics.


\section{Conclusion}

In this paper, we propose a voting scheme with post-quantum security based the physical laws. The advantages of our scheme are as follows. First, the post-quantum security of the voting scheme is based on the physical laws, which depends on the inherent limitations of quantum computers. Due to the inherent limitations, the security is not influenced by the evolution of new quantum algorithms. Second, some generic cryptographic components, which consist of the lightweight password, the symmetric algorithm and the unconditional secure MAC, are applied in the scheme making the scheme easier to implement in practice. Third, the scheme is based on the quantum computer physical limitation reaching all the post-quantum security properties. Finally, we emphasize the point that the presented scheme based on the viewpoint -``active defense" is a superior and productive research direction. we emphasize that the use of the practical post-quantum security based on physical laws to achieve a voting scheme should be an attractive and fruitful research approach. We look forward to further research in this direction of work.

\end{document}